\documentstyle[aps,12pt,preprint]{revtex}
\begin{document}
\preprint{SU-ITP-93-14}
\title{Quasi-Fermi  Distribution and Resonant Tunneling of Quasiparticles
with Fractional Charges}
\author{V.L. Pokrovsky}
\address{Physics Dept., Texas A\&M University, College
        Stat., TX 77843\cite{PrAddr}\\
   {\em and}\\L.D. Landau Institute for Theoretical Physics,\\
   V-334, Kosygina 2, Moscow, GSP-1, 117940, Russia}
\author{L.P. Pryadko}
\address{Physics Dept., Stanford University, Stanford, CA
94305\cite{Email}}
\date{May 18, 1993}
\maketitle
\begin{abstract}
We study the resonant tunneling of quasiparticles through an impurity
between the edges of a Fractional Quantum Hall sample. We show that the
one-particle momentum distribution of fractionally charged edge
quasiparticles has a quasi-Fermi character. The density of states near the
quasi-Fermi energy at zero temperature is singular due to  the
statistical interaction of quasiparticles. Another effect of this
interaction is a new selection rule for the resonant tunneling of
fractionally charged quasiparticles: the resonance is suppressed unless
an integer number of {\em electrons} occupies the impurity. It allows
a new explanation of the scaling behavior observed in the mesoscopic
fluctuations of the conductivity in the FQHE.
\end{abstract}

\

The question regarding the extent to which the fractionally charged
quasiparticles proposed by
Laughlin~\cite{Laughlin} are real and whether they can be observed
individually, was recently
resolved by experiments~\cite{Exper}. In these experiments FQHE
samples with constriction were studied, in order to observe
essentially one-particle tunneling processes of the quasiparticles. In
particular, the frequencies of the mesoscopic fluctuations of the longitudinal
resistance in the FQHE with $\nu=1/3$ were compared to those in the
Integer QHE.

Previously, Jain and Kivelson~\cite{Jain-Kivel} suggested that the the
resonant tunneling of electrons from one edge to another through an
impurity could cause an enhancement of the dissipative resistance in the
IQHE samples with a narrow  constriction. Kivelson and
Pokrovsky~\cite{Kivel-Pokr} proposed an analogous mechanism for the
fractionally charged quasiparticles in the FQHE. Their model implied
simple scaling laws for the periods of the mesoscopic oscillations in the
vicinity of the state with filling factor $\nu\equiv p/q\!:$ $\Delta
B\propto q$ at fixed gate voltage $V_G,$  and $\Delta V_G\propto p$ at
fixed magnetic field $B$. Both predictions have been confirmed
experimentally~\cite{Exper}.

Nevertheless, the theoretical understanding of this scaling can not be
considered as satisfactory. In particular, Kivelson~\cite{Kivelson:Semi}
derived quasiclassical quantization rules for a multi-anyon bound state
at the impurity allowing for the statistical interaction; his scaling
relations are different from the observed ones. P. Lee~\cite{lee}
supported Kivelson's result from the position of the theory of edge
quasiparticles. He accounted for the obvious discrepancy with the
experiment with the Coulomb blockade.

Another problem in understanding these experiments is that the mechanism
of resonant tunneling usually implicates the existence of a Fermi level
for excitations. It clearly exists for the case of the IQHE but is much
less obvious for the FQHE. Recently Haldane~\cite{Haldane} defined
the generalized Pauli principle for anyons. This principle, however, does
not imply the existence of the distinct Fermi level required to explain
the resonances in tunneling.

The purpose of this work is to elucidate these general questions and give
a new explanation for the experimental result.

Let us start with the quasiclassical quantization rule derived by
Kivelson. It reads
        \begin{equation} \Phi=m\phi^*_0+N\phi_0, \label{flux}
        \end{equation}
where $\Phi$ is the total magnetic flux through an area $A$ surrounded by
the trajectory of the quasiparticle, $\phi_0=hc/e$ is the flux quantum for
an electron, $\phi^*_0=q\phi_0$ is the flux quantum for a quasiparticle
(anyon) with charge $e^*=e/q$, $N$ is the number of quasiparticles
captured by the impurity, and $m$ is the angular momentum of the tunneling
quasiparticle. The first term in the r.h.s. of equation~(\ref{flux}) is
required by the gauge invariance, while the second one simply shows that
each quasiparticle is bound with one flux quantum. The same quantum
spectrum arises in the exact solution for a system of $N$ anyons in a
quadratic potential~\cite{exact}.

At a given gate voltage $V_G,$ the area $A$ enclosed by the trajectory,
corresponding to the Fermi level, is the same for any quantized value of
Hall conductivity. This is true because the Laughlin liquid is
incompressible. Therefore, the intervals of the magnetic field between
consequent bound states of a quasiparticle are $\Delta B_q=\phi^*_0/A$ if
the number of quasiparticles $N$ is fixed. The scaling, consistent with
these intervals, was observed experimentally.

However, during the tunneling the number of quasiparticles $N$ coupled
with the impurity changes by one.  It corresponds to the change of the
flux $\Phi$ by a single flux quantum $\phi_0$ instead of
$\phi_0^*=q\phi_0.$ Corresponding periods $\Delta B_1=\phi_0/A$ have {\em
not} been observed experimentally.

The solution to this puzzle lies in the fractional statistics of
quasiparticles. Consider the situation where $N$ quasiparticles are
initially bound to the impurity, and the tunneling quasiparticle arrives
at an orbit enclosing all of them. In the quasiclassical approximation,
the wave function of this quasiparticle will gain a phase factor
$z=exp(i 2\pi N/q)$ after each complete revolution around the quantized
orbit.  More accurately, it is multiplied by $z(1-\gamma/2),$ where
$\gamma$ is the total probability of tunneling from the
impurity to either left or right edge. The total tunneling amplitude
contains a series
        \begin{equation} \label{quasiclassic}
        t_{LR}=\sum_k z^k(1-\gamma/2)^k=\frac{1}{1-z(1-\gamma/2)}.
        \end{equation}
Usually resonant enhancement of the tunneling happens when all the
amplitudes, corresponding to different numbers of revolutions in
(\ref{quasiclassic}) are coherent, \mbox{\em i.e.} $z=1.$ This is
obviously the case for the usual  Fermi quasiparticles ($\nu=1$). For a
fractional value of $\nu,$ the contributions of $q$ consequent revolutions
almost cancel each other. Thus the {\em resonant} tunneling is suppressed
unless $N/q$ is an integer. In other words, the tunneling of an anyon is
resonantly enhanced only if an integer number of electrons are already
bound to the impurity. This simple selection rule restores the scaling
suggested in ref.~\cite{Kivel-Pokr} and agrees with experiment.

The scaling of the oscillation intervals on the gate voltage $\Delta
V_G$~\cite{Kivel-Pokr} is also easily reproduced. Indeed, at a fixed
magnetic field $B$ the change $\Delta V_G$ corresponding to a new
resonance is determined by the change of the area
        \begin{equation} \Delta A=\Delta\Phi/B,
        \end{equation}
where $\Delta\Phi$ is the change of the flux through the trajectory. As we
have already established, $\Delta\Phi=q\phi_0$ for the resonant tunneling
at $\nu=p/q.$ On the other hand, the value of the magnetic field
$B_{\nu},$ corresponding to the filling factor $\nu,$ is approximately
$1/\nu$ times $B_{1}.$ As a result we obtain $\Delta A_{\nu}=p\Delta
A_{1}$ and $\Delta V_G^{\nu}=p\Delta V_G^1$

These intuitive and semi-classical arguments are supported by direct
calculations in the framework of Wen's theory of edge
excitations~\cite{Wen-r}. Simultaneously, we find the distribution of edge
quasiparticles over momenta to confirm the conjecture of its Fermi-like
character~\cite{Kivel-Pokr}. All calculations have been performed for
special values of $\nu=1/q,$ where q is an odd integer.

In Wen's theory the operator creating a quasiparticle
        \begin{equation}\psi^{\dag}(x,t)= \,\,:\!e^{i\phi(x,t)}\!:
        \label{anyon}
        \end{equation}
at the point $x$ of the edge is associated with the chiral
Bose field $\phi(x,t)$ of an edge magneto-plasmon. This field obeys the
commutation relationship
        \begin{equation}  [\phi_{x,t},\phi_{x',t}]=- i \pi\nu
        \,\text{sign}(x-x')
        \end{equation}
and is related to the charge density
$\rho={e}/{2\pi}\,{\partial\phi}/{\partial x}$ at the edge.
The permutation relations of the anyon operators~(\ref{anyon}) are
        \begin{equation} \psi^{\dag}(x,t)\psi^{\dag}(x',t)=e^{ i
\pi\nu\,\text{sign}(x-x')}
             \psi^{\dag}(x',t)\psi^{\dag}(x,t). \label{permutation}
        \end{equation}
To find the distribution of edge quasiparticles over momenta, it is
necessary to calculate the Fourier-transformation $\tilde{G}_p$ of the
simultaneous correlation function
$G(x-x')=\langle\psi^{\dag}(x,t)\psi(x',t)\rangle.$ We have performed this
calculation explicitly with the following result: $\tilde{G}_p$ can be
represented as the product
        \begin{equation}\label{split}
        \tilde{G}(p')=g_{_T}(p')\frac{1}{\exp(\beta p' v)+1},
        \end{equation}
where the momentum $p'=p-p_F,$ $p_F$ is a Fermi momentum, and $v$ is
the drift velocity along the edge for both the chiral field and the anyons.
The second factor is the usual Fermi-distribution, while the first one can
be treated
as the temperature-dependent density of states; it is an even function of
$p'.$ At $T=0$ the density of states $g_{_T}(p')$ has a singularity
$\propto |p'|^{\nu-1}$ and diverges at the Fermi-level. The singularity is
smeared out at a finite temperature. Details of the calculations will be
published elsewhere.

To investigate the tunneling, we modelled the impurity as a void in the
incompressible quantum Hall liquid, its edge being an additional
environment for the edge quasiparticles. The perimeter $L_i$ of this edge
was assumed to be small enough to neglect the probability of thermal
excitation of states with non-zero angular momentum $m.$ The outer edge,
on the contrary, was assumed to be in the thermodynamical limit; it
serves as a thermostat. The many-body quantum mechanical states at the
impurity are well-defined in the limit of a small tunneling coupling
        \begin{equation} H_t=\int dx dy t(x,y) \psi^{\dag}(x)
            \psi_i(y)+h.c.,
        \end{equation}
where $\psi_i(y)$ is the annihilation operator for the edge
quasiparticles at the impurity. This limit allows us to reduce the evolution
equation of the density matrix at the impurity to a set of kinetic
equations, describing the evolution of probabilities
$W_N=\langle{\hat{\cal P}}_N\rangle$ to have exactly $N$ quasiparticles at
the impurity, where ${\hat{\cal P}}_N$ is the appropriate projection
operator.

The transition amplitudes are connected with different equilibrium averages
similar
to $\langle\psi_i^{\dag}(y)\psi_i(y'){\hat{\cal P}}_N\rangle$. In the
generic case they are periodic only at the $q$-fold
boundary~\cite{Endnote}, gaining the phase $2\pi N/q$ in each cycle. This
phase, similar to the Berry phase, is the exact
consequence of the fractional statistics and does not depend on the
distribution of edge phonons. As usual, the broken symmetry leads to the
selection rule for the allowed transitions; namely, transitions
$N\longrightarrow N+1$ are suppressed unless $N/q$ is integer. This
statement coincides with our conclusion extracted from the semi-classical
model. In contrast to the previous derivation, we made no assumptions
about the geometrical properties of the orbit of the tunneling
quasiparticle.

The factorization~(\ref{split}) yields the Boltzmann distribution for the
probabilities of different many-particle states of the impurity in
equilibrium. In the presence of the inter-edge potential difference,
however, the impurity population depends on the tunneling probabilities.
The current-voltage dependence (see Figure~\ref{picture}) is highly
non-linear and asymmetric, especially in the  vicinity of the resonance.

In conclusion, we have found new selection rules for the resonant
tunneling of quasiparticles in the FQHE, arising from the broken symmetry
specific to anyons. The equilibrium momentum distribution of the edge
quasiparticles has quasi-Fermi properties with the temperature-dependent
density of states. This explains the appearance of  resonant tunneling
effects in the anyonic systems and, in particular, the scaling properties
of the mesoscopic pattern measured in the experiment~\cite{Exper}.

Upon completion of this work, we received the
preprint~\cite{preprint}, where the RG equations for
resonant inter-edge tunneling are solved numerically in a different
geometry. The authors considered neither scaling properties of the
resonant effects, nor the momentum distribution of quasiparticles. Their
main emphasis was the line shape.

V. L. P. is indebted to Steve Kivelson for numerous discussions and
to J. S. Langer and Institute of Theoretical Physics in Santa-Barbara
for the hospitality extended to him at the initial stage of this work.
L. P. P. wishes to thank the Soros Foundation for partial financial
support under grant \# S92.56, and Jared Levy for valuable comments on
the manuscript.

\begin{figure}
\caption{ Non-linear resonant tunneling current $I$ versus the inter-edge
voltage $V$ expressed in the units of the temperature $T$ at different
values of the one-particle energy $E$ of the bound state at the impurity.}
\label{picture}
\end{figure}

\begin{references}
\bibitem{Endnote}
Our definition of the normal ordering is
different from the conventional one in the treatment of the zero momentum
mode of the chiral field $\phi.$
\end{references}
\end{document}